\documentclass{elsart3}

\usepackage{amssymb}
\usepackage{amsmath}
\usepackage{graphicx}

\begin{document}

\begin{frontmatter}

  \title{Phase-field simulations of solidification in binary and
    ternary systems using a finite element method}

  \author[IAF]{D.~Danilov}
  \ead{denis.danilov@fh-karlsruhe.de}
  \and
  \author[CS]{B.~Nestler}
  \ead{britta.nestler@fh-karlsruhe.de}

  \address[IAF]{Institute of Applied Research, Karlsruhe University of
    Applied Sciences, Moltkestrasse 30, D-76133 Karlsruhe Germany}

  \address[CS]{Department of Computer Science, Karlsruhe University of
    Applied Sciences, Moltkestrasse 30, D-76133 Karlsruhe Germany}

  \begin{abstract}
    We present adaptive finite element simulations of dendritic and
    eutectic solidification in binary and ternary alloys.  The
    computations are based on a recently formulated phase-field model
    that is especially appropriate for modelling non-isothermal
    solidification in multicomponent multiphase systems. In this
    approach, a set of governing equations for the phase-field
    variables, for the concentrations of the alloy components and for
    the temperature has to be solved numerically, ensuring local
    entropy production and the conservation of mass and inner energy.
    To efficiently perform numerical simulations, we developed a
    numerical scheme to solve the governing equations using a finite
    element method on an adaptive non-uniform mesh with highest
    resolution in the regions of the phase boundaries.  Simulation
    results of the solidification in ternary
    Ni$_{60}$Cu$_{40-x}$Cr$_{x}$ alloys are presented investigating
    the influence of the alloy composition on the growth morphology
    and on the growth velocity. A morphology diagram is obtained that
    shows a transition from a dendritic to a globular structure with
    increasing Cr concentrations.  Furthermore, we comment on 2D and
    3D simulations of binary eutectic phase transformations. Regular
    oscillatory growth structures are observed combined with a
    topological change of the matrix phase in 3D. An outlook for the
    application of our methods to describe AlCu eutectics is given.
  \end{abstract}


\end{frontmatter}

\section{Introduction}
Multicomponent alloys form the most important class of metallic
materials for technical and industrial processes. Combined with the
number of components is a wealth of different phases, phase
transformations, complex thermodynamic interactions and pattern
formations. The complex phase diagrams of multicomponent alloys show a
variety of different solidification processes such as eutectic,
peritectic and monotectic types of reaction. Since the most common
solidification microstructures occuring in industrial alloys, such as
Al and Fe based alloys, are dendrites and eutectic composite
structures, we will concentrate our investigations on these two types
of structures.

The phase-field modelling technique has made a significant progress
during the last decade in simulating complex microstructures, such as
dendritic and eutectic growth; see e.g. \cite{NW2,BWBK}.  The newly
introduced variable $\phi(x,t)$ in these models indicates the physical
state of a system at each point in space.  Differing from sharp
interface concepts, $\phi({x},t)$ has a smooth transition at the
interfaces (phase boundaries). In the bulk phases, $\phi({x},t)$ takes
on constant values, e.g. $\phi({x},t) = 0$ for the liquid and
$\phi({x},t)=1$ for the solid phase. The governing equations consist
of parabolic partial differential equations for the evolution of the
phase state coupled to mass and heat transport equations. They serve
for performing simulations of complex growth morphologies without
explicitly tracking the phase boundaries.  In a recent article
\cite{GarckeSIAM}, the phase-field methodology has been generalized to
the case of arbitrary numbers of $N$ components and $M$ phases in
alloy systems by introducing a vector of concentrations
${c}=(c_1,\ldots,c_N)$ and a vector of phase-field variables
${\phi}=(\phi_1,\ldots,\phi_M)$.

For computations based on diffuse interface models, it is required
that the spatial resolution of the numerical method must be greater
than the thickness of the diffusive phase boundary layer. The
interfacial thickness itself must be less than the characteristic
scale of the growing microstructure. In this case, a non-uniform grid
with adaptive refinement can dramatically reduce the use of
computational resources against a uniform grid with the same spatial
resolution.  For these reasons, we use an adaptive finite element
method for solving the governing equations of the new multicomponent
phase-field model.

The present report is part of our intension to apply the phase-field
model and the numerical method to simulate solidification processes in
real alloys which are usually multicomponent.  We present simulation
results of ternary dendritic and of binary eutectic growth.  In
particular, the ternary Ni$_{60}$Cu$_{40-x}$Cr$_{x}$ alloy system is
considered to investigate the influence of interplaying solute fields
on the interface stability, on the growth velocity and on the
characteristic type of morphology. By choosing the Ni-Cu-Cr system, we
build upon the binary Ni-Cu system which has been explored by
phase-field modelling (e.g. \cite{Warren1995}) and by molecular
dynamics simulations in several papers (e.g. \cite{Hoyt}).  Hence,
physical parameters are relatively well established.  For our
application to eutectic solidification, we constructed a symmetric
binary model phase diagram. Special emphasis is laid on a discussion
of oscillatory growth structures.

\section{Phase-field model}

The phase-field model in \cite{GarckeSIAM} has been formulated in a
thermodynamically consistent way and allows for an arbitrary number of
phases and components. It is defined solely via bulk free energies of
the individual phases, the surface energy densities of the interfaces,
the diffusion and mobility coefficients and yields classical moving
boundary problems in the sharp interface limit. The approach is based
on an entropy density functional $S(e, c, \phi)$ of the form
\begin{equation}
  \label{eq:entropy_functional}
  S = \int_{\Omega} \left[
    s(e, c, \phi)
    - \left(
      \varepsilon a(\phi, \nabla \phi)
      + \frac{1}{\varepsilon} w(\phi)
    \right)
    \right] dx.
\end{equation}
The bulk entropy density $s$ depends on the internal energy $e$, on
the concentrations of components ${c}=(c_i)_{i=1}^{N}$, and on the
phase-field variable ${\phi}=(\phi_{\alpha})_{\alpha=1}^{M}$.  The
thermodynamics of the interfaces is given by the second and third term
in the integral of Eq.~(\ref{eq:entropy_functional}) and is determined
by the gradient energy density $a(\phi, \nabla\phi)$, the multi well
potential $w(\phi)$ and a small scale parameter $\varepsilon$ related
to the thickness of the interface. The gradient energy and the multi
well potential depend on the surface energy density $\sigma$ and its
anisotropy can be taken into account using appropriate choices of
$a(\phi, \nabla\phi)$.  The variable $\phi_{\alpha}$ with
$0\leqslant\phi_{\alpha}\leqslant 1$ denotes the local fraction of
phase $\alpha$. It is required that the concentrations and phase-field
variables fulfill the constraints
\begin{equation}
  \label{eq:constraints}
  \sum_{i=1}^{N} c_i = 1,
  \quad
  \sum_{\alpha=1}^{M} \phi_{\alpha} = 1.
\end{equation}
The evolution equations for the phase fields are postulated as
\begin{equation}
  \label{eq:phase-field}
  \omega\varepsilon \partial_{t} \phi_\alpha
  = \varepsilon
  ( \nabla a_{,\nabla \phi_\alpha} - a_{,\phi_\alpha} )
  - \frac{1}{\varepsilon} w_{,\phi_\alpha}
  - \frac{f_{,\phi_\alpha}}{T}
  - \lambda,
\end{equation}
where $f(T,{c},{\phi})$ is the free energy density and $\lambda$ is a
Lagrange multiplier such that the constraint in
Eq.~(\ref{eq:constraints}) for phase fields is satisfied. The kinetic
factor $\omega =\omega(\phi, \nabla\phi)$ describes anisotropic
interface kinetics and is related to the kinetic coefficient $\mu$ of
atomic attachment in the linear response function ``growth
velocity--interfacial undercooling'' at a flat front.

Considering an ideal solution system, we define the free energy
density as follows
\begin{equation}
  \label{eq:free-energy-density}
  f(c,\phi)
  = \sum_{i=1}^N \sum_{\alpha=1}^M
  c_i L_i^{\alpha} \frac{T-T_i^{\alpha}}{T_i^{\alpha}}
  h(\phi_{\alpha})
  + \frac{RT}{v_m}\sum_{i=1}^N c_i \ln c_i,
\end{equation}
where $L_i^{\alpha}$ and $T_i^{\alpha}$ are the latent heat and the
melting temperature of the component $i$ in the phase $\alpha$,
respectively. $R$ is the gas constant, $v_m$ is the molar volume and
$h(\phi_{\alpha})$ is a monotone function on the interval $[0,1]$
satisfying $h(0)=0$ and $h(1)=1$. Additionally, we consider an
isothermal approximation with the temperature $T$ being constant.

Assuming that the mass fluxes are linear functions of the
thermodynamic driving forces, i.e. chemical potentials in isothermal
approximation, mass balance equations can be written as
\begin{equation}
  \label{eq:mass-balance}
  \partial_{t} c_i
  = - \nabla \left(
  \sum_{j=1}^N
  L_{ij}(c,\phi) \nabla \frac{-\mu_j}{T}
  \right),
\end{equation}
with chemical potentials $\mu_i = f_{,c_i}$ and mobility coefficients
given by
\begin{equation}
  \label{eq:mobility-coefficients}
  L_{ij}(c,\phi) = \frac{v_m}{R}
  D_i c_i
  \left(
    \delta_{ij} - \frac{D_j c_j}{\sum_{k=1}^3 D_k c_k}
  \right).
\end{equation}
The form of Eq. (\ref{eq:mobility-coefficients}) allows different
values of the bare trace diffusion coefficients $D_i(\phi)$ for the
different components $i$ and satisfies the constraint for
concentrations in Eq.~(\ref{eq:constraints}).

The evolution Eqns.~(\ref{eq:phase-field}) and (\ref{eq:mass-balance})
are solved using a finite element method with Lagrange elements and
linear test functions. For the time evolution, a semi-implicit
formulation is discretized to achieve better numerical stability.  We
generate a non-uniform adaptive mesh having the highest order of
spatial resolution in the vicinity of the solid--liquid interface
where the gradients of phase fields and concentrations reach maximal
values. The mesh structure is adopted in time according to the
evolution of phase and concentration fields using a refinement
criterion based on the phase-field and concentration gradients.

\section{Dendritic growth in ternary alloys}

By numerical simulations, we investigate how the change of the alloy
composition influences the growth morphology for a given
solidification condition.  In particular, we fix the initial
undercooling and choose a ternary Ni$_{60}$Cu$_{40-x}$Cr$_{x}$ alloy
as a prototype for this study. To recover the corresponding solidus
and liquidus lines of the binary Ni-Cu and Ni-Cr phase diagrams in the
region of concentrations up to 40\,at.\% of Cu or Cr, the following
physical parameters are used: Melting temperatures
$T_{\mathrm{Ni}}=1728$~K, $T_{\mathrm{Cu}}=1358$~K,
$\widetilde{T}_{\mathrm{Cr}}=1465$~K; latent heats
$L_{\mathrm{Ni}}=2350$~J/cm$^3$, $L_{\mathrm{Cu}}=1728$~J/cm$^3$,
$\widetilde{L}_{\mathrm{Cr}}=1493$~J/cm$^3$ and a molar volume
$v_m=7.42$~cm$^3$. Since we do not consider the complete eutectic
Ni-Cr phase diagram, the values of the melting temperature and of the
latent heat for Cr are not the real physical data, but adjustable
parameters in order to recover the actual binary phase diagram in the
given region of concentrations. They are marked by a tilde to
emphasize this difference from the data for Ni and Cu.  Applying these
values leads to a partition coefficient $k_e=0.843$, to a liquidus
slope $m_e=-3.27$~K/at.\%, and to a freezing range $\Delta
T_0=m_e(k_e-1)/k_e c_\infty=24$~K for binary Ni$_{60}$Cu$_{40}$.
Similarly, we obtain $k_e=0.905$, $m_e=-2.08$~K/at.\%, and $\Delta
T_0=8.7$~K for the binary Ni$_{60}$Cr$_{40}$ system and the
corresponding equilibrium phase diagram.

We assume that both surface properties, the surface energy density
$\sigma$ and the kinetic coefficient $\mu$ do not depend on the alloy
composition and have the values $\sigma=0.37$~J/m$^2$ and
$\mu=3.3$~mm/(s K), refering to \cite{Warren1995}. The anisotropy of
the interface properties plays an important role in the selection of
the operating state during dendritic growth. In this study, we use the
values calculated from molecular dynamics simulations. In \cite{Hoyt},
the strength of the surface energy anisotropy is given as $0.023$ and
the strength for kinetic anisotropy as $0.169$.

The diffusion coefficients in the melt are $D_{\mathrm{Ni}}=3.82\times
10^{-9}$~m$^2$/s \cite{Hoyt}, $D_{\mathrm{Cu}}=3.32\times
10^{-9}$~m$^2$/s \cite{Hoyt} and $D_{\mathrm{Cr}}=1.5\times
10^{-9}$~m$^2$/s. The diffusion coefficients in the solid phase are
set as an equal value $10^{-13}$~m$^2$/s for all components.  The
small length scale parameter $\varepsilon$ in the entropy functional
(Eq.~(\ref{eq:entropy_functional})) is $\varepsilon=0.1$~$\mu$m.

\begin{figure}[b]
  \centering
  \includegraphics[width=0.47\textwidth]{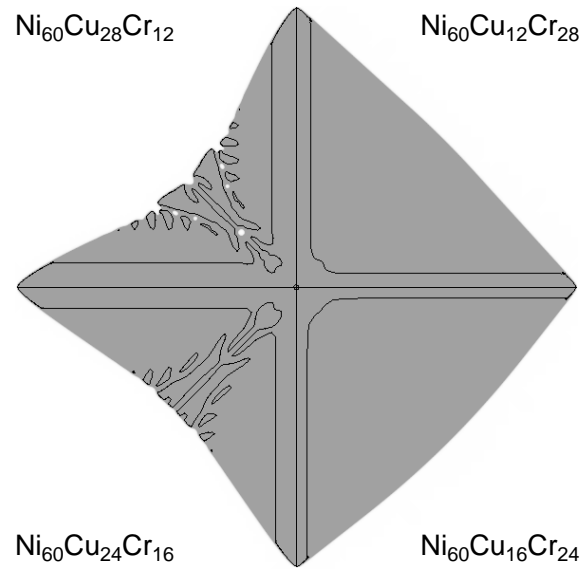}
   \caption{\label{fig:struct}
     Dendritic to globular morphological transition for different
     alloy compositions. The atomic percents of Cu and Cr are
     exchanged keeping Ni fixed at 60~at.\%. The shaded regions
     correspond to the solid phase and the solid lines represent the
     isolines of average concentration of Ni in the solid phase.  }
\end{figure}

Using the physical parameters given above, we carried out a series of
numerical computations for different alloy compositions varying from
Ni$_{60}$Cu$_{36}$Cr$_{4}$ to Ni$_{60}$Cu$_{4}$Cr$_{36}$. The
concentration of Ni was kept at 60~at.\% and the initial undercooling
was fixed at $20$~K measured from the equilibrium liquidus line in the
phase diagram at a given composition of the melt. A morphological
transition from dendritic to globular growth occurs at a melt
composition of about Ni$_{60}$Cu$_{20}$Cr$_{20}$. The left side of
Fig.~\ref{fig:struct} shows the dendritic morphologies observed for Cr
concentrations less than 20~at.\%. The right side of the figure
displays globular morphologies for Cr concentration crossing this
threshold.  The velocity of the dendritic/globular tip increases
linearly from 1.19~cm/s to 3.24~cm/s with increasing the concentration
of Cr.  An analogous type of dendritic to globular morphology
transition depending on the undercooling has been observed in
numerical simulations of binary alloy solidification and is discussed
in \cite{Galenko}.

Qualitatively, such transitions can be expected by comparing the
initial undercooling $\Delta T$ with the freezing range $\Delta T_0$:
For the binary Ni$_{60}$Cu$_{40}$ alloy, the undercooling $\Delta T=
20$~K is less than the freezing range $\Delta T_0=24$~K, whereas for
the binary Ni$_{60}$Cr$_{40}$ system the undercooling $\Delta T=20$~K
is greater than $\Delta T_0=8.7$~K.  Thus the morphology change is
related to the transition from a two-phase region (above the solidus
line) to a one-phase region (below the solidus) in correspondance with
the phase diagram. Qualitative analysis of growth structures should
take into account the kinetic effects of the solid-liquid interface as
well as the surface tension acting as stabilizing factor and the
diffusion properties of alloy components as destabilizing forces due
to concentration gradients \cite{KF}.

\section{Simulation of binary eutectic growth in 2D and 3D}

\begin{figure}
  \centering
  \includegraphics[width=0.47\textwidth]{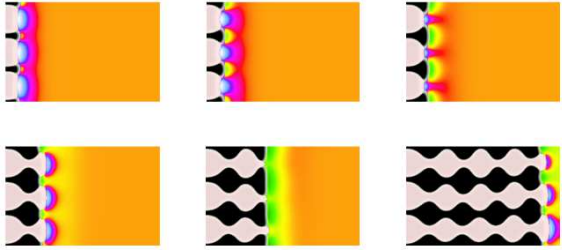}
  \caption{\label{osci}
    Regular oscillations along the solid-solid interface during binary
    eutectic growth driven by the motion of the triple junction in 2D.
  }
\end{figure}

To simulate eutectic phase transitions in a binary A--B alloy, where
two solid phases $\alpha$ and $\beta$ grow into an undercooled melt,
we have constructed a typical eutectic phase diagram via the method of
common tangents. In our computations, we consider the widely observed
phenomenon of regular oscillations along the solid-solid interface
driven by the triple junctions (Fig. \ref{osci}).  The initial
concentration of the melt is chosen at the eutectic composition
$c_{A}=c_{B}=0.5$, where the two solid phases grow with equal phase
fractions. A transition from steady state lamellar growth to an
oscillatory pattern formation is found for an increasing difference of
the initially set up phase fraction.  For a ratio of 1:3 between the
initial phase fraction of solid $\alpha$:$\beta$, lamellar growth is
re-established. For ratios of 1:4 and 1:6, regular oscillations with a
characteristic amplitude and wave length as in Fig. (\ref{osci}) can
be observed, whereas for a ratio 1:7, the $\beta$ phase overgrows the
initially dominating $\alpha$ solid phase. In such a case, a
nucleation event takes place.  Two-dimensional oscillatory structures
have been discussed in \cite{Karma96} by boundary integral method and
more recently in \cite{Kim04} by means of phase-field modelling.

\begin{figure}
  \centering
  \includegraphics[width=0.47\textwidth]{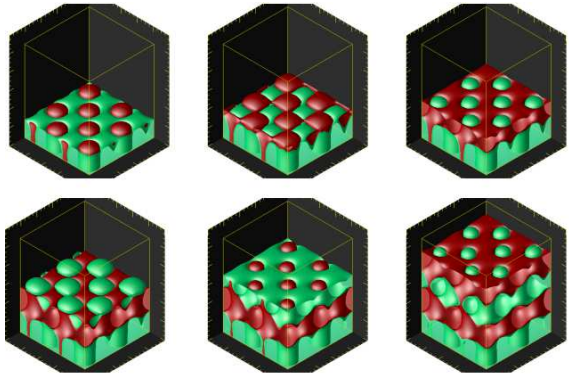}
  \caption{\label{3Dosci}
    Topological change of the microstructure with $\alpha $ rods
    embedded in a $\beta $ matrix phase and visa versa.  The formation
    results from regular 3D oscillations along the solid-solid
    interface.  }
\end{figure}

The analogous type of oscillation can be observed for eutectic
microstructure formations in three dimensions, Fig. \ref{3Dosci}.
Performing an alternating topological change, $\alpha $ solid rods are
embedded in a $\beta $ matrix followed by the opposite situation of
$\beta $ crystals embedded in an $\alpha $ matrix. Further
three-dimensional simulations of eutectic microstructures are reported
in \cite{Apel02,Lewis04}

\section{Conclusion}
A new phase-field model for non-isothermal solidification in
multicomponent multiphase alloy systems has been applied to ternary
Ni-Cu-Cr primary phase growth and to binary eutectic growth of two
solid phases into an undercooled melt. A transition from a dendritic
to a globular structure is observed in our simulation series, while
successively exchanging Cu by Cr. The experiences with a binary model
eutectic will enable us to apply the modelling and simulation
technique in eutectic and off-eutectic Al-Cu alloys in a forthcoming
paper. Steady-state lamellar domains as well as different types of
oscillations persisting over a wide range of compositions and growth
rates are reported in \cite{ZKC}.

The authors gratefully acknowledge helpful discussions with Mathis
Plapp and the financial support provided by the German Research
Foundation (DFG) under Grant No. Ne~882/2.


\end{document}